\documentclass[aps,prl,twocolumn,showpacs,preprintnumbers,amsmath,amssymb]{revtex4-1}

\usepackage{graphicx} % Include figure files

\begin{document}

\preprint{}

\title{Three-wave mixing with three incoming waves: Signal-Idler Coherent Cancellation and Gain Enhancement in a Parametric Amplifier}

\author{Flavius Schackert}
  \email{flavius.schackert@yale.edu}
\author{Ananda Roy}
\author{Michael Hatridge}
\author{A. Douglas Stone}
\author{Michel H. Devoret}
\affiliation{Departments of Physics and Applied Physics, Yale University, 15 Prospect Street, New Haven, CT 06511}

%\date{\today}
\date{December 9, 2012}

\begin{abstract}
Coherent, purely-dispersive three-wave mixing systems in optics and superconducting microwave circuits can be operated as parametric amplifiers, generating from a pump wave at one frequency amplified signal and idler waves at lower frequencies. Here we demonstrate the reciprocal process using a Josephson amplifier in which coherently imposed signal and idler beams up-convert to the pump frequency. For signal and idler beams strong enough to significantly deplete the pump, we show that this reciprocal process (``coherent cancellation'') leads to large, phase-sensitive modulation and even enhancement of the amplifier gain, in good agreement with
theoretical predictions. 
\end{abstract}

\pacs{42.65.Ky, 42.65.Yj, 85.25.Cp, 85.25.Dq}

\maketitle
Parametric amplification based on three-wave mixing is a fundamental process in electromagnetic signal processing \cite{Franken_PRL1961}, both in the optical and microwave frequency domain. More recently, with the advent of quantum information science, three-wave mixing provides a basic building block for measurements at the single photon level  \cite{Aspect_PRL1981, Kwiat_PRL1995}, where it is crucial that the nonlinear mixing process is purely dispersive. An important class of parametric amplifiers make use of three-wave mixing to amplify incoming signal fields through down-conversion of a higher frequency pump field. The amplification process involves incoming pump photons at angular frequency $\omega_P$ being split up into outgoing signal and idler photons at frequencies $\omega_S$ and $\omega_I$ respectively, where $\omega_P = \omega_S+\omega_I$. The three-wave mixing equations for the photon fields leads, under the undepleted (stiff) pump approximation, to a linear two-port scattering matrix for the signal and idler fields \cite{Louisell_Book1960}.       
As recently emphasized \cite{Longhi_PRL2011}, the symmetry of the three-wave mixing equations at the classical level implies that the parametric process can be operated in reverse, converting signal and idler photons, in presence of the pump, into additional pump photons. We refer to this reversed process as coherent cancellation (CC). Unlike the typical amplification process, in which only signal and pump beams are present as inputs, coherent cancellation requires three coherent input beams: along with the pump, both signal and idler must be present and balanced in amplitude, and there must be a specific phase relation of the three beams. In this case, and in contrast to a matched termination where the power is absorbed and converted into heat, all of the incident signal and idler power undergoes CC and reappears at the pump port. Hence, the pump oscillation inside the device is enhanced by signal and idler, and the gain can actually increase beyond its undepleted value.  
%As the relative phase of the three beams is varied away from the CC condition, for relatively strong signal and idler inputs, the gain of the amplifier will oscillate periodically 

The Hamiltonian of a three-wave mixing device \cite{Bergeal_NatPhys2010, Dykman_Book2012}, under the rotating wave approximation (RWA), neglecting external drive and signal fields, is 
\begin{equation}
H^\mathrm{RWA}=\hbar\omega_a a^\dagger a+\hbar\omega_b b^\dagger b+\hbar\omega_c c^\dagger c+\hbar g_3(a^\dagger b^\dagger c+ab c^\dagger),
\label{eq:Hamiltonian}
\end{equation}
where $a$, $b$, and $c$ are the annihilation operators of the signal, idler, and pump modes of center frequency $\omega_a$, $\omega_b$, $\omega_c$, and bandwidth $\kappa_a$, $\kappa_b$, $\kappa_c$, respectively, and  $g_3$ is the coupling strength. The term $a^\dagger b^\dagger c$, exploited for amplification (annihilation of a pump photon for the creation of a pair of signal and idler photons), is accompanied by its counter-part $ab c^\dagger$, which describes the new operation that can be seen as the reciprocal of amplification (annihilation of one signal and idler photon together leading to the creation of a pump photon). The work we present here reveals this process demanded by the time-reversal symmetry of the Hamiltonian, (\ref{eq:Hamiltonian}). We observe this both by measuring the attenuation of the signal and idler beams when their relative phase is tuned to the coherent cancellation condition, and, more directly, by observing gain increase at the CC point by its effect on an additional probe tone.  The latter effect is very difficult to observe in almost all practical amplifiers. Practical devices are designed with parameters optimized for gain, bandwidth, and stability, resulting in vastly different scattering properties of pump and signal/idler modes. Signal and idler powers and bandwidths are typically several orders of magnitude lower than those of the pump, so that subtle effects in the pump dynamics are hidden under a large background field, and thus not observable in lossy three-wave mixing systems. 
%The pump presents a large quasi-constant coherent tone, and its dynamics are usually undetectable. (Isn't this a bit redundant?)
However, with the advent of superconducting Josephson amplifiers operating at the quantum limit \cite{Castellanos_NatPhys2008,Bergeal_Nature2010,Hatridge_PRB2011,Roch_PRL2012}, we possess sufficient control of all relevant degrees of freedom to observe reverse parametric effects.  The CC effect and gain enhancement demonstrated in this paper are semi-classical in nature, but the effects could also be observed in the full quantum regime, where zero-point fluctuations of the fields would dominate.

Our device is a widely tunable Josephson Parametric Converter (JPC) \cite{Roch_PRL2012, Hatridge_Science2012}, which can be operated as a non-degenerate phase-preserving parametric amplifier \cite{Bergeal_Nature2010, Abdo_APL2011} or as a noiseless frequency converter \cite{Abdo_InPreparation2012}, at microwave frequencies. 
\begin{figure}[h]
  \begin{center}
    \includegraphics[scale=1]{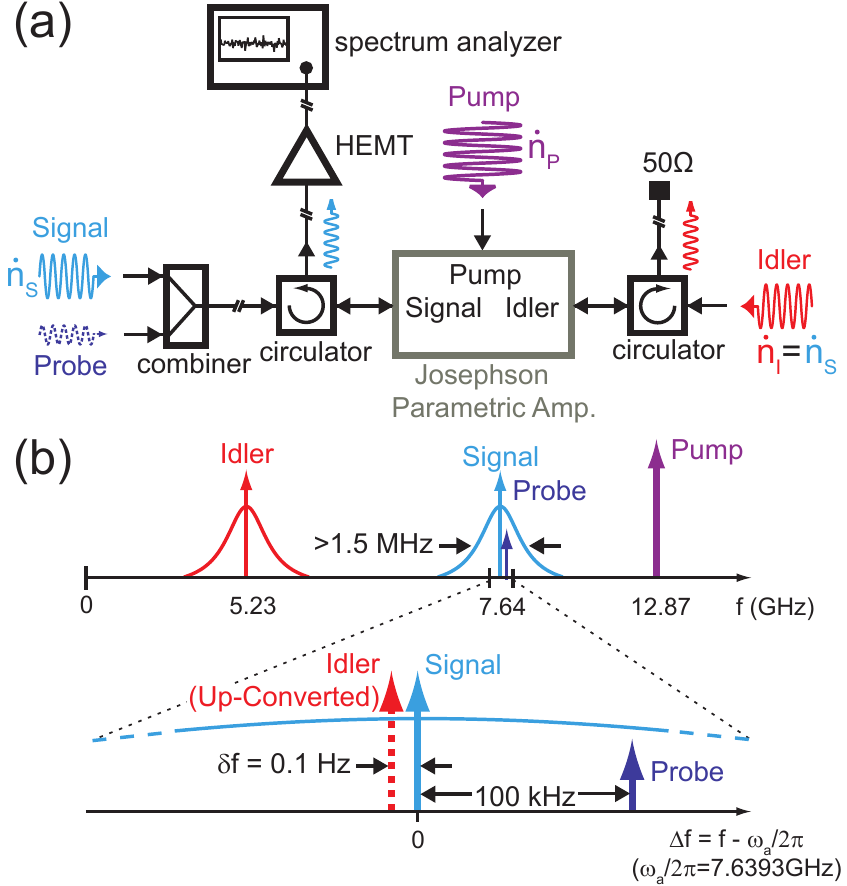}
    \caption{(a) Schematic representation of the measurement setup. Four continuous wave (CW) tones are injected in the three ports of the superconducting Josephson parametric amplifier (gray box). Incoming signal, idler, and pump photon fluxes are designated by $\dot{n}_S$, $\dot{n}_I$, and $\dot{n}_P$, respectively. The output power from the signal port is measured with a spectrum analyzer (SA).
(b) Upper: Applied CW tones in frequency space. The horizontal axis is the frequency axis, the colored arrows represent coherent tones, and the Lorentzian shapes represent the gain response function. Lower: enlargement of the signal band. The dashed arrow represents the up-converted idler tone, offset by $\delta f=0.1$ Hz from the signal tone, producing a slow variation of the relative phase $\phi=2\pi\delta f\cdot t$ between signal and idler tones.
}
    \label{fig:1} %Label must come before end of center, otherwise labeling will refer to an incorrect number, but after \caption{}
  \end{center}
\end{figure}
The JPC, operated at $30$ mK in a cryogen-free dilution refrigerator, has three ports (Fig. \ref{fig:1}(a)), one each to access the signal (centered around $\omega_a/2\pi= 7.6393$ GHz), idler (centered around $\omega_b/2\pi=5.2277$ GHz), and pump (at $\omega_c/2\pi=12.867$ GHz) modes. Cryogenic circulators on the signal and idler ports separate input and output waves, allowing them to travel on different transmission lines (see \cite{Bergeal_Nature2010} and \cite{Abdo_APL2011} for details on the setup). Four phase-locked microwave generators provide tones to the signal port (signal and probe tones), the idler port (idler tone), and the pump port (pump tone). Incoming signal, idler, and pump photon fluxes are designated by $\dot{n}_S$, $\dot{n}_I$, and $\dot{n}_P$, respectively. We monitor in time the JPC signal port output power through a spectrum analyzer (SA) set to zero span mode, i.e. set to a frequency window given by the spectrum analyzer's resolution bandwidth (RBW) and center frequency. 
The additional small amplitude probe tone used to measure gain modulation is offset from the JPC signal center frequency by $100$ kHz (Fig. \ref{fig:1}(b)), which allows its detection with the SA without contamination from the large amplitude signal tone at the center of the JPC signal band, as long as the RBW of the SA is set to be much smaller than $100$ kHz. Both the signal and the probe tone are well within the amplification bandwidth of the JPC, which is $1.5$ MHz at a gain of $25$ dB, and which increases at smaller gains according to the gain bandwidth product $\sqrt{G_0}B=\pi^{-1}/({\kappa_a}^{-1}+{\kappa_b}^{-1})$ \cite{Bergeal_NatPhys2010}, where $G_0$ is the power gain and $B$ the dynamical amplification bandwidth (in Hz).

The coherent cancellation effect is observed simply by injecting coherent signal and idler tones (with no probe tone at this stage) with the correct relative phase so that they destructively interfere, leading to a strong attenuation of the corresponding output tones by a factor determined by the gain of the amplifier.
The effect is described by the linear two-port scattering matrix of the undepleted pump approximation:
From the reduced JPC scattering matrix \cite{Bergeal_NatPhys2010}
\begin{equation}
 S=\begin{pmatrix}\sqrt{G_0}&-ie^{-i\varphi_p}\sqrt{G_0-1}\\ie^{i\varphi_p}\sqrt{G_0-1}&\sqrt{G_0}\end{pmatrix},
 \label{eq:matrix}
\end{equation}
where $G_0$ is the (undepleted) gain of the amplifier when only a weak signal tone is present and  $\varphi_P$ is the pump phase.
This phase will be kept fixed in the subsequent analysis and can be set equal to $\pi/2$ without loss of generality, although it 
should be emphasized that CC is a {\it three wave}, nonlinear interference effect, and can be modulated with the pump phase 
as well as the relative phase of signal and idler.
The eigenvalues of this S-matrix are reciprocals: 
\begin{math}
\lambda_+ = \sqrt{G_0} + \sqrt{G_0-1} = 2 \sqrt{G_0} + \varepsilon , \ \varepsilon = o(G_0) \ \mathrm{for} \ G_0 \to \infty; \ \lambda_- =1/ \lambda_+= \sqrt{G_0} - \sqrt{G_0-1} = (2 \sqrt{G_0}+\varepsilon)^{-1}
\end{math}, 
the former corresponding to power amplification $4G_0$ and the latter to power attenuation $(4G_0)^{-1}$ at large gains.  If $A$ is the signal amplitude (corresponding to photon flux 
$\dot{n}_S = |A|^2$), the eigenvectors are
$\vec{e}_+ = (A,-A),\vec{e}_- = (A,A)$, requiring a balanced idler input ($\dot{n}_I = \dot{n}_S$), either in phase or $\pi$ out of phase.  The usual situation, when only the signal tone is sent in, corresponds to an equal superposition of the two eigenvectors, and only the amplification mode can be observed.  If balanced beams with relative phase 
$\phi$ are input, the  normalized output power, $P=P_{\mathrm{out}}/(G_0 P_{\mathrm{in}})$, (where $P_{\mathrm{out}},P_{\mathrm{in}}$ are the signal output and input power) is given by
\begin{equation}
\label{eq:cc}
P=2-G_0^{-1}-2\sqrt{1-G_0^{-1}}\cos{\phi},
\end{equation} 
varying between increased gain 
$(P \approx 4)$ and strong attenuation ($P \approx (4G_0^2)^{-1}$).  As the amplifier approaches the oscillation threshold, $G_0 \to \infty$ and the 
cancellation becomes perfect \cite{Longhi_PRL2011}.  This effect is similar to the time-reversed laser operation in a coherent perfect absorber (CPA) \cite{Chong_PRL2010, Wan_Science2011}. Contrary to a CPA however, here the two input beams have different frequencies, and instead of absorption to an unspecified dissipative sink, here the input photons are converted into pump photons at their sum-frequency.  

%We demonstrate in this letter the reverse parametric process in a superconducting Josephson amplifier by observing the depletion and enhancement of the amplifier gain, which is directly related to the pump field amplitude inside the device. This variation of gain is measured by injecting a fourth weak probe tone on the signal port. The phase-sensitive depletion and enhancement of the pump field manifests itself when applying two continuous wave (CW) signal and idler input tones with carefully balanced incoming photon fluxes ($\dot{n}_S=\dot{n}_I$), while also applying a strong incoming CW pump tone. By monitoring the signal port reflection coefficient as a function of the phase $\phi$ between signal and idler input tones, we observe the coherent cancellation/enhancement of signal power as a function of $\phi$, with the predicted dependence on the system gain. %%%%%%%%%%%%%%%%%%%%%%%%%%%%%%%%%%%%%%%%%%%%%%%%%%%%%%%%%%%%%%%%%%%%%%%%%%%%%%%%%%%%%%%%%%%%%
%%%%%%								Experimental Setup and Figure 1								
%%%%%%%%%%%%%%%%%%%%%%%%%%%%%%%%%%%%%%%%%%%%%%%%%%%%%%%%%%%%%%%%%%%%%%%%%%%%%%%%%%%%%%%%%%%%%

To observe CC in the JPC we slowly vary the phase $\phi$ between signal and idler input tones at a rate of $0.1$ Hz, by offsetting the idler tone  
above the JPC idler mode center frequency by $\delta f=0.1$ Hz. The parametric amplification process up-converts and amplifies the idler tone, which, as a result of the detuning, appears $\delta f$ below the amplified signal tone. To be able to monitor the power at the signal port in time with sufficient resolution of the phase $\phi=2\pi\delta f\cdot t$, we set the SA to a RBW that is faster then the detuning $\delta f$: $36$ Hz when observing the attenuation of the signal tone and $51$ Hz when monitoring the probe gain \footnote{We chose to present these two data sets as this is where we best achieved the balance between signal and idler photon flux.}(see Supplemental Material at [URL will be inserted by publisher] for choice of bandwidths and detunings).  
%%%%%%%%%%%%%%%%%%%%%%%%%%%%%%%%%%%%%%%%%%%%%%
%%%%%%% 	MOVE TO SUPPLEMENT   %%%%%%%%%%%%%%%
%%%%%%%%%%%%%%%%%%%%%%%%%%%%%%%%%%%%%%%%%%%%%%
%These RBW's are fast enough to capture the dynamics of the JPC as a function of $\phi$, while being small enough for the experiment to have a sufficient signal-to-noise ratio (SNR), and also making sure not to capture the signal/idler tones when measuring the probe tone power. We would like to emphasize the hierarchy of all relevant bandwidths and frequency offsets: The largest bandwidth is the JPC dynamical bandwidth, which is $\geq 1.5$ MHz, followed by the probe tone offset of $100$ kHz (still within the JPC dynamical bandwidth). Then comes the SA RBW of at least $36$ Hz (captures either signal/idler or probe tone power, but never both) and finally the detuning $\delta f=0.1$ Hz (much smaller than RBW to have enough phase resolution).    
%%%%%%%%%%%%%%%%%%%%%%%%%%%%%%%%%%%%%%%%%%%%%%%%%%%%%%%%%%%%%%%%%%%%%%%%%%%%%%%%%%%%%%%%%%%%%%%%%%%%%%%%%%%%%%%%%%%%%%%%
%%%%%%%%%%%%%%%%%%%%									Coherent Cancellation and Figure 2								%%%%%%%%%%%%%%%%%%%%%%%%%%%%%%%%%%%%%%%%%%%%%%%%%%%%%%%%%%%%%%%%%%%%%%%%%%%%%%%%%%%%%%%%%%%%%%%%%%%%%%%%%%%%%%%%%%%%%%%%
\begin{figure}[h]
  \begin{center}
    \includegraphics[scale=1]{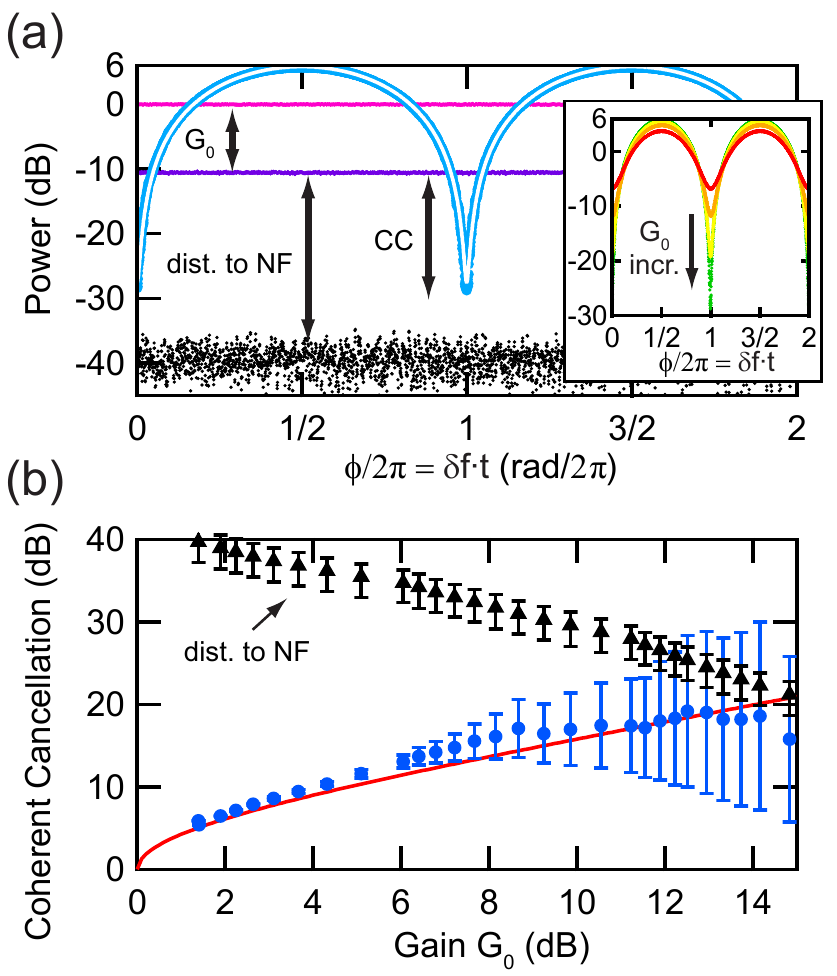}
    \caption{(a) Normalized output power as a function of $\phi$ (light blue) and fit to Eq. \ref{eq:cc} (white), for balanced signal and idler incoming photon fluxes ($\dot{n}_S=\dot{n}_I$) and $G_0=11$ dB. Calibration traces are as follows: idler turned off (pink); idler and pump off (purple); signal and idler off (i.e. noise floor, black). Arrows indicate how various quantities (gain ($G_0$), coherent cancellation (CC), distance of zero cancellation to noise floor (dist. to NF)) are measured. Inset: shown traces correspond to gains varying from $1$ (red) to $11$ dB (green).
(b) Magnitude of CC versus gain at maximum interference point (blue). Red solid trace represents the theory, and the black triangles the resolution limitations due to the finite distance to the noise floor. 
}
    \label{fig:2} %Label must come before end of center, otherwise labeling will refer to an incorrect number, but after \caption{}
  \end{center}
\end{figure}
Fig. \ref{fig:2} shows the CC of the input signal tone, for signal and idler input powers well below (at least $10$ dB) the $1$ dB amplifier compression point \cite{Pozar_Book1997}. The light blue trace is the measured, normalized signal output power $P$, as function of the relative phase $\phi$ between signal and idler. In addition, the 
data is fit to Eq. \ref{eq:cc} which is then plotted as the overlaid white curve; the single fit parameter, $G_0 = 11$ dB agrees well with its independently measured
value. As noted, at large gains, and $\phi=(2n+1)\pi$, $n\in \mathbb{Z}$, the wave amplitudes interfere constructively to produce a normalized power output $6$ dB (factor of $4$) above the signal output power with idler tone off ($G_0 P_{\mathrm{in}}$, pink trace). CC manifests itself at relative phase $\phi=2\pi n$, when only a fraction $(4G_0)^{-1}$ of the incident power leaves the JPC through the signal and idler ports (normalized power $P=(4G_0^2)^{-1}$).
%Quantitatively, at large gains, incoming signal fields are amplified by a factor 
%$4G$ at the constructive interference point and are attenuated by the same factor $4G$ at the destructive interference (coherent cancellation, labeled CC in Fig. \ref{fig:2}(a)) point.       
The data in the inset in Fig. \ref{fig:2}(a) shows $P$ for increasing $G_0$, confirming this behavior.
    
The coherent cancellation is a measure of the efficiency of the conversion of signal and idler photons into pump photons. Fig. \ref{fig:2}(b) shows that there is good agreement between our data (blue dots) and the prediction of Eq. \ref{eq:cc} (red) for all gains up to the experimental limit imposed by the system noise floor (black triangles; see arrow labeled `dist. to NF' in Fig. \ref{fig:2}(a)). All error bars are calculated from the noise floor data according to the Dicke radiometer formula \cite{Dicke_RevSciInstrum1946}. The increase for the CC data at larger gains is due to the fact that we systematically decrease the signal and idler tone powers at larger gains to make sure to stay well below the saturation point of the device (we keep $G_0\cdot \dot{n}_S$ approximately constant while ensuring that we exactly have $\dot{n}_S=\dot{n}_I$). This leads to a decrease in signal-to-noise ratio (SNR), as the JPC noise (and thus the system noise) increases with $G_0$. A more serious limitation to the CC measurement is the fact that the noise floor (black trace in Fig. \ref{fig:2}(a)) is pushed up when $G_0$ increases, while the input signal is adjusted to decrease. This means that our ability to observe the CC effect decreases for increasing gains (see `dist. to NF' in Fig. \ref{fig:2}(b)). This could of course be improved by reducing the RBW of the spectrum analyzer (pushing down the noise floor), but at the expense of having to perform the experiments slower by the same factor to keep the current phase resolution, which would make it more sensitive to 1/f noise and phase drifts.
%%%%%%%%%%%%%%%%%%%%%%%%%%%%%%%%%%%%%%%%%%%%%%%%%%%%%%%%%%%%%%%%%%%%%%%%%%%%%%%%%%%%%%%%%%%%%%%%%%%%%%
%%%%%%%%%%%%%%%%%%%%%%%%%%%%     Gain Modulation and Figure 3
%%%%%%%%%%%%%%%%%%%%%%%%%%%%%%%%%%%%%%%%%%%%%%%%%%%%%%%%%%%%%%%%%%%%%%%%%%%%%%%%%%%%%%%%%%%%%%%%%%%%%%

The coherent cancellation effect as described by Eq. \ref{eq:cc} is easily calculated in the undepleted pump approximation, but it is more general and
still applies in the regime of larger signal and idler relative to the pump, when the depletion or enhancement of the pump is significant. The gain modulation effects in this regime provide a method to confirm that the attenuated signal and idler photons at the CC condition are not lost to some other dissipative process, but are being coherently converted to pump photons. Specifically, as the relative phase, $\phi$ of the signal and idler is varied, and the JPC is alternatively amplifying and attenuating these inputs, the pump is either 
depleted or strengthened, and this can be observed as a phase dependent modulation of the effective gain experienced by the weak probe tone, which we now introduce.

Experimentally, we keep the signal and idler photon fluxes equal ($\dot{n}_S=\dot{n}_I$) but in contrast to the experiment of Fig.
\ref{fig:2}, with signal and idler amplitudes large enough to significantly saturate the device. Eq. \ref{eq:cc} still holds, with the undepleted gain, $G_0$ replaced by a nonlinear gain, $G=G(\phi,x)$, $x=\dot{n}_S/\dot{n}_P$, which must be calculated self-consistently.
\begin{figure}[h]
  \begin{center}
    \includegraphics[scale=1]{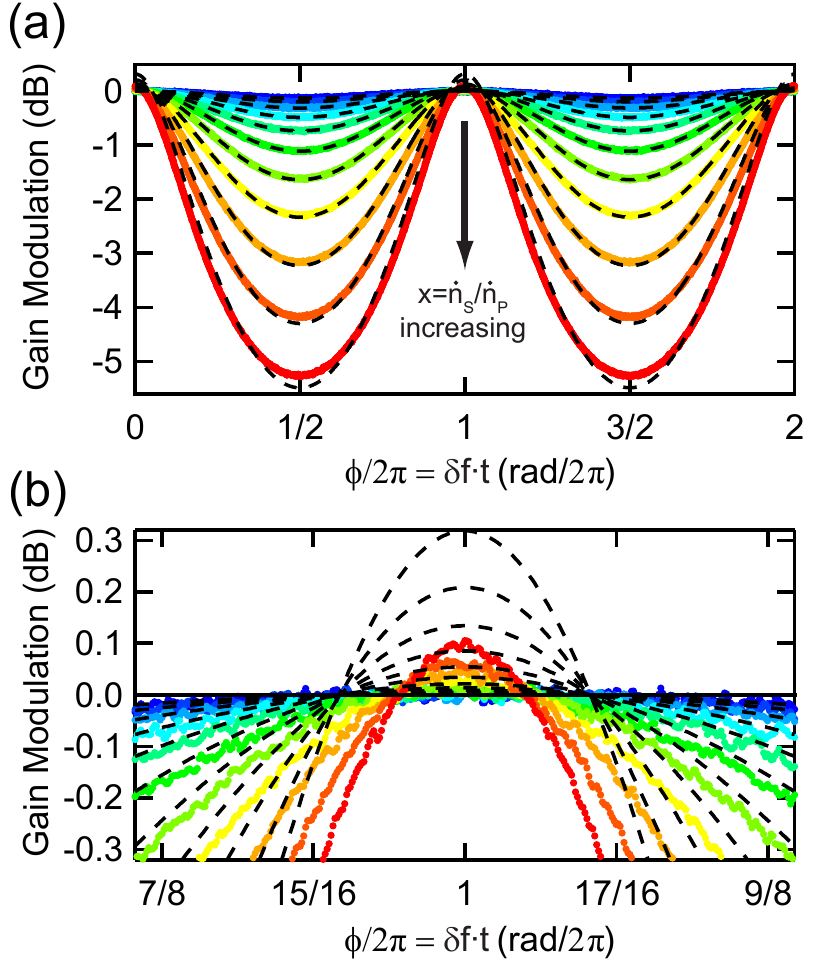}
    \caption{(a) Phase-dependent depletion of signal port gain ($G_0=11$ dB). Amplified probe power is measured in a $51$ Hz RBW (signal tone is offset by $100$ kHz). Traces correspond to signal and idler photon fluxes increasing by $2$ dB (from blue to red). Dashed lines are least square fits of gain modulation theory.
(b) Enlargement of the region around $\phi = 2\pi$. Gain enhancement above $G_0=11$ dB ($0$ dB line) is clearly visible for experimental data. 
}
    \label{fig:3} %Label must come before end of center, otherwise labeling will refer to an incorrect number, but after \caption{}
  \end{center}
\end{figure}
Fig. \ref{fig:3}(a) shows the expected large modulation of the gain with $\phi$, and, most dramatically, around $\phi=2\pi$ (see Fig. \ref{fig:3}(b)), we observe an {\it increase} in the JPC gain; moreover, the gain enhancement increases as we increase $x$ in $2$ dB steps  (colors from blue to red). The data set shown corresponds to an undepleted gain of $11$ dB (normalized to $0$ dB line), and an applied pump power that is not changed across traces. Note that the gain enhancement  
$\sim$$x$, whereas the gain depletion $\sim$$4 G_0 x$, so one expects the enhancement to be smaller than the depletion.

We can calculate $G=G(\phi,x)$ by solving the equations of motion derived from the three-wave mixing Hamiltonian (Eq. \ref{eq:Hamiltonian}) iteratively in the correction to the undepleted pump approximation (see Supplemental Material at [URL will be inserted by publisher] for details on derivation). This leads to the self-consistent equations for the pump parameter $\rho\in[0,1)$:
\begin{eqnarray}
 \rho&=&\rho_0\Big|\Big[1-i\frac{\rho_0}{4}x\frac{1}{\sqrt{G}}\Big\{
(1+\sqrt {G})^2e^{i(\phi+\varphi_p)}\nonumber\\&&-2i(1+\sqrt{G})\sqrt{G-1}-e^{
-i(\phi+\varphi_p) }
(G-1)\Big\}\Big]\Big|,\\\sqrt{G}&=&\frac{1+\rho^2}{1-\rho^2},
\end{eqnarray}where $\rho_0=\sqrt{(G_0-1)/(G_0+1)}\in[0,1)$ is the undepleted pump parameter, and $G_0$ is the undepleted gain (no signal/idler inputs). 
%Note that the gain depletion/enhancement cannot be predicted by the (non-unitary) signal-idler scattering matrix of Eq. \ref{eq:matrix}, as only the amplifier behavior in the stiff pump regime is described by it.
The dashed lines in Fig. \ref{fig:3} correspond to theory with a single fit parameter $x$. As expected, the fits reproduce the $2$ dB steps of applied signal and idler powers (not shown).   
We find excellent agreement between our theory and our experiment for phases $\phi$ away from $2\pi n$, where signal and idler fields constructively interfere and lead to a depletion of the pump photon flux, manifested as a decrease of the JPC gain. 
%%%%%%%%%%%%%%%%%%%%%%%%%%%%%%%%%%%%%%%%%%%%%%%%%%%%%%%%%%%%%%%%%%%%%%%%%%%%%%%%%%%%%%%%%%%%%%%%%%%%%%%%%%%%%%%%%%%%%%%%			Figure 4
%%%%%%%%%%%%%%%%%%%%%%%%%%%%%%%%%%%%%%%%%%%%%%%%%%%%%%%%%%%%%%%%%%%%%%%%%%%%%%%%%%%%%%%%%%%%%%%%%%%%% 
\begin{figure}[h]
  \begin{center}
    \includegraphics[scale=1]{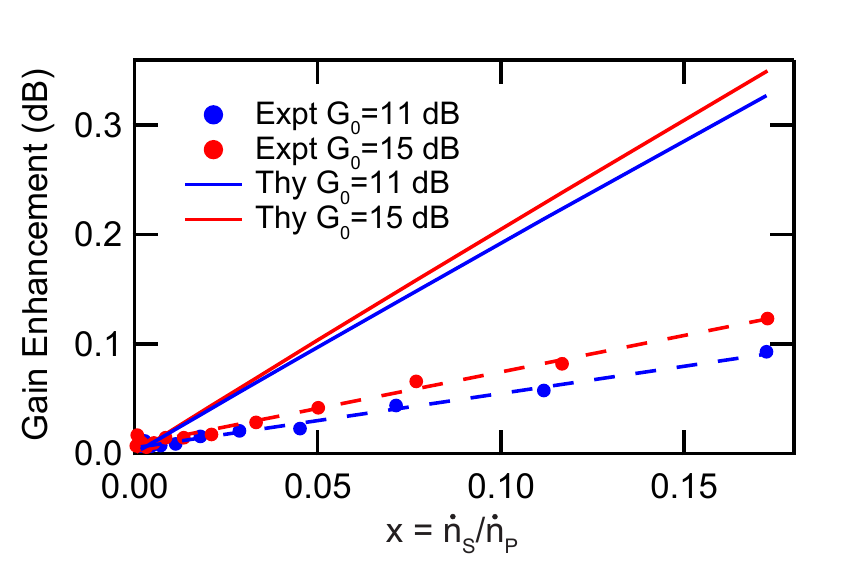}
    \caption{Experimental gain enhancement data (circles) and theory (lines) for $G_0=11$ dB (blue) and $G_0=15$ dB (red). The gain enhancement is evaluated at a relative phase $\phi$ between signal and idler tones of $2\pi$ and plotted against photon flux ratio $x$.
}
    \label{fig:4} %Label must come before end of center, otherwise labeling will refer to an incorrect number, but after \caption{}
  \end{center}
\end{figure}

Figure \ref{fig:4} shows a direct comparison of the experimental gain enhancement to the theoretical prediction at $\phi=2\pi n$ as a function of $x=\dot{n}_S/\dot{n}_P$ for gains of $11$ dB and $15$ dB. Although the experimental slope is about $1/3$ of the expected theoretical slope, we unambiguously observe significant gain enhancement in our JPC, the key signature of the coherent conversion of signal and idler photons into pump photons.
There are several reasons for the discrepancy between experimental and theoretical gain enhancement: the experiment requires significant averaging with the given bandwidths and powers (ultimately limited by the JPC dynamical bandwidth and dynamic range), while each modulation is rather slow ($0.1$ Hz). It is thus sensitive to 1/f noise and microwave generator phase drifts. Further, any mismatch between $\dot{n}_S$ and $\dot{n}_I$ decreases the gain enhancement. Finally, there may be a contribution due to spurious dissipation in the CC process. 
%%%%%%%%%%%%%%%%%%%%%%%%%%%%%%%%%%%%%%%%%%%%%%%%%%%%%%%%%%%%%%%%%%%%%%%%%%%%%%%%%%%%%%%%%%%%%%%%%%%%%%%%%%%%%%
%%%%%%%%%%%% Conclusion
%%%%%%%%%%%%%%%%%%%%%%%%%%%%%%%%%%%%%%%%%%%%%%%%%%%%%%%%%%%%%%%%%%%%%%%%%%%%%%%%%%%%%%%%%%%%%%%%%%%%%%%%%%%%%%

The presented gain enhancement results can be understood as a benchmark for the level of control we have over all degrees of freedom in our three-wave mixing device. Any hidden (uncontrolled) degree of freedom other than the signal, idler, and pump modes will inevitably be perceived as dissipation and thus lead to a reduction in the gain enhancement effect.  A potential application of the reverse operation of a JPC is its use in phase-locking two coherent tones of different frequencies in a feedback loop, which could be crucial for quantum information processing with artificial atoms.

%%%%%%%%%%%%%%%%%%%%%%%%%%%%%%%%%%%%%%%%%%%%%%%%%%%%%%%%%%%%%%%%%%%%%%%%%%%%%%%%%%%%%%%%%%%%%%%%%%%%%
%%%%%%%%%%%%		Acknowledgements:
%%%%%%%%%%%%%%%%%%%%%%%%%%%%%%%%%%%%%%%%%%%%%%%%%%%%%%%%%%%%%%%%%%%%%%%%%%%%%%%%%%%%%%%%%%%%%%%%%%%%%
Discussions with Hui Cao are gratefully acknowledged. 
This research was supported by IARPA under grant W911NF-09-1-0369, ARO under grant W911NF-09-1-0514 and NSF under grants DMR-1006060 and DMR-0653377. A. D. S. acknowledges support by NSF under grant ECCS-1068642. Facilities used were supported by Yale Institute for Nanoscience and Quantum Engineering and NSF MRSEC DMR 1119826.
  
%\bibliography{references}{}

\begin{thebibliography}{19}%
\makeatletter
\providecommand \@ifxundefined [1]{%
 \@ifx{#1\undefined}
}%
\providecommand \@ifnum [1]{%
 \ifnum #1\expandafter \@firstoftwo
 \else \expandafter \@secondoftwo
 \fi
}%
\providecommand \@ifx [1]{%
 \ifx #1\expandafter \@firstoftwo
 \else \expandafter \@secondoftwo
 \fi
}%
\providecommand \natexlab [1]{#1}%
\providecommand \enquote  [1]{``#1''}%
\providecommand \bibnamefont  [1]{#1}%
\providecommand \bibfnamefont [1]{#1}%
\providecommand \citenamefont [1]{#1}%
\providecommand \href@noop [0]{\@secondoftwo}%
\providecommand \href [0]{\begingroup \@sanitize@url \@href}%
\providecommand \@href[1]{\@@startlink{#1}\@@href}%
\providecommand \@@href[1]{\endgroup#1\@@endlink}%
\providecommand \@sanitize@url [0]{\catcode `\\12\catcode `\$12\catcode
  `\&12\catcode `\#12\catcode `\^12\catcode `\_12\catcode `\%12\relax}%
\providecommand \@@startlink[1]{}%
\providecommand \@@endlink[0]{}%
\providecommand \url  [0]{\begingroup\@sanitize@url \@url }%
\providecommand \@url [1]{\endgroup\@href {#1}{\urlprefix }}%
\providecommand \urlprefix  [0]{URL }%
\providecommand \Eprint [0]{\href }%
\providecommand \doibase [0]{http://dx.doi.org/}%
\providecommand \selectlanguage [0]{\@gobble}%
\providecommand \bibinfo  [0]{\@secondoftwo}%
\providecommand \bibfield  [0]{\@secondoftwo}%
\providecommand \translation [1]{[#1]}%
\providecommand \BibitemOpen [0]{}%
\providecommand \bibitemStop [0]{}%
\providecommand \bibitemNoStop [0]{.\EOS\space}%
\providecommand \EOS [0]{\spacefactor3000\relax}%
\providecommand \BibitemShut  [1]{\csname bibitem#1\endcsname}%
\let\auto@bib@innerbib\@empty
%</preamble>
\bibitem [{\citenamefont {Franken}\ \emph {et~al.}(1961)\citenamefont
  {Franken}, \citenamefont {Hill}, \citenamefont {Peters},\ and\ \citenamefont
  {Weinreich}}]{Franken_PRL1961}%
  \BibitemOpen
  \bibfield  {author} {\bibinfo {author} {\bibfnamefont {P.~A.}\ \bibnamefont
  {Franken}}, \bibinfo {author} {\bibfnamefont {A.~E.}\ \bibnamefont {Hill}},
  \bibinfo {author} {\bibfnamefont {C.~W.}\ \bibnamefont {Peters}}, \ and\
  \bibinfo {author} {\bibfnamefont {G.}~\bibnamefont {Weinreich}},\ }\href
  {\doibase 10.1103/PhysRevLett.7.118} {\bibfield  {journal} {\bibinfo
  {journal} {Phys. Rev. Lett.}\ }\textbf {\bibinfo {volume} {7}},\ \bibinfo
  {pages} {118} (\bibinfo {year} {1961})}\BibitemShut {NoStop}%
\bibitem [{\citenamefont {Aspect}\ \emph {et~al.}(1981)\citenamefont {Aspect},
  \citenamefont {Grangier},\ and\ \citenamefont {Roger}}]{Aspect_PRL1981}%
  \BibitemOpen
  \bibfield  {author} {\bibinfo {author} {\bibfnamefont {A.}~\bibnamefont
  {Aspect}}, \bibinfo {author} {\bibfnamefont {P.}~\bibnamefont {Grangier}}, \
  and\ \bibinfo {author} {\bibfnamefont {G.}~\bibnamefont {Roger}},\ }\href
  {\doibase 10.1103/PhysRevLett.47.460} {\bibfield  {journal} {\bibinfo
  {journal} {Phys. Rev. Lett.}\ }\textbf {\bibinfo {volume} {47}},\ \bibinfo
  {pages} {460} (\bibinfo {year} {1981})}\BibitemShut {NoStop}%
\bibitem [{\citenamefont {Kwiat}\ \emph {et~al.}(1995)\citenamefont {Kwiat},
  \citenamefont {Mattle}, \citenamefont {Weinfurter}, \citenamefont
  {Zeilinger}, \citenamefont {Sergienko},\ and\ \citenamefont
  {Shih}}]{Kwiat_PRL1995}%
  \BibitemOpen
  \bibfield  {author} {\bibinfo {author} {\bibfnamefont {P.~G.}\ \bibnamefont
  {Kwiat}}, \bibinfo {author} {\bibfnamefont {K.}~\bibnamefont {Mattle}},
  \bibinfo {author} {\bibfnamefont {H.}~\bibnamefont {Weinfurter}}, \bibinfo
  {author} {\bibfnamefont {A.}~\bibnamefont {Zeilinger}}, \bibinfo {author}
  {\bibfnamefont {A.~V.}\ \bibnamefont {Sergienko}}, \ and\ \bibinfo {author}
  {\bibfnamefont {Y.}~\bibnamefont {Shih}},\ }\href {\doibase
  10.1103/PhysRevLett.75.4337} {\bibfield  {journal} {\bibinfo  {journal}
  {Phys. Rev. Lett.}\ }\textbf {\bibinfo {volume} {75}},\ \bibinfo {pages}
  {4337} (\bibinfo {year} {1995})}\BibitemShut {NoStop}%
\bibitem [{\citenamefont {Louisell}(1960)}]{Louisell_Book1960}%
  \BibitemOpen
  \bibfield  {author} {\bibinfo {author} {\bibfnamefont {W.}~\bibnamefont
  {Louisell}},\ }\href@noop {} {\emph {\bibinfo {title} {Coupled mode and
  parametric electronics}}}\ (\bibinfo  {publisher} {Wiley},\ \bibinfo {year}
  {1960})\BibitemShut {NoStop}%
\bibitem [{\citenamefont {Longhi}(2011)}]{Longhi_PRL2011}%
  \BibitemOpen
  \bibfield  {author} {\bibinfo {author} {\bibfnamefont {S.}~\bibnamefont
  {Longhi}},\ }\href {\doibase 10.1103/PhysRevLett.107.033901} {\bibfield
  {journal} {\bibinfo  {journal} {Phys. Rev. Lett.}\ }\textbf {\bibinfo
  {volume} {107}},\ \bibinfo {pages} {033901} (\bibinfo {year}
  {2011})}\BibitemShut {NoStop}%
\bibitem [{\citenamefont {Bergeal}\ \emph
  {et~al.}(2010{\natexlab{a}})\citenamefont {Bergeal}, \citenamefont {Vijay},
  \citenamefont {Manucharyan}, \citenamefont {Siddiqi}, \citenamefont
  {Schoelkopf}, \citenamefont {Girvin},\ and\ \citenamefont
  {Devoret}}]{Bergeal_NatPhys2010}%
  \BibitemOpen
  \bibfield  {author} {\bibinfo {author} {\bibfnamefont {N.}~\bibnamefont
  {Bergeal}}, \bibinfo {author} {\bibfnamefont {R.}~\bibnamefont {Vijay}},
  \bibinfo {author} {\bibfnamefont {V.~E.}\ \bibnamefont {Manucharyan}},
  \bibinfo {author} {\bibfnamefont {I.}~\bibnamefont {Siddiqi}}, \bibinfo
  {author} {\bibfnamefont {R.~J.}\ \bibnamefont {Schoelkopf}}, \bibinfo
  {author} {\bibfnamefont {S.~M.}\ \bibnamefont {Girvin}}, \ and\ \bibinfo
  {author} {\bibfnamefont {M.~H.}\ \bibnamefont {Devoret}},\ }\href
  {http://dx.doi.org/10.1038/nphys1516} {\bibfield  {journal} {\bibinfo
  {journal} {Nat. Phys.}\ }\textbf {\bibinfo {volume} {6}},\ \bibinfo {pages}
  {296} (\bibinfo {year} {2010}{\natexlab{a}})}\BibitemShut {NoStop}%
\bibitem [{\citenamefont {Abdo}\ \emph
  {et~al.}(2012{\natexlab{a}})\citenamefont {Abdo}, \citenamefont {Kamal},\
  and\ \citenamefont {Devoret}}]{Dykman_Book2012}%
  \BibitemOpen
  \bibfield  {author} {\bibinfo {author} {\bibfnamefont {B.}~\bibnamefont
  {Abdo}}, \bibinfo {author} {\bibfnamefont {A.}~\bibnamefont {Kamal}}, \ and\
  \bibinfo {author} {\bibfnamefont {M.~H.}\ \bibnamefont {Devoret}},\ }\href
  {http://books.google.com/books?id=tj0uVYnXLcUC} {\emph {\bibinfo {title}
  {Fluctuating Nonlinear Oscillators: From Nanomechanics to Quantum
  Superconducting Circuits}}},\ edited by\ \bibinfo {editor} {\bibfnamefont
  {M.}~\bibnamefont {Dykman}}\ (\bibinfo  {publisher} {OUP Oxford},\ \bibinfo
  {year} {2012})\ pp.\ \bibinfo {pages} {119--141}\BibitemShut {NoStop}%
\bibitem [{\citenamefont {Castellanos-Beltran}\ \emph
  {et~al.}(2008)\citenamefont {Castellanos-Beltran}, \citenamefont {Irwin},
  \citenamefont {Hilton}, \citenamefont {Vale},\ and\ \citenamefont
  {Lehnert}}]{Castellanos_NatPhys2008}%
  \BibitemOpen
  \bibfield  {author} {\bibinfo {author} {\bibfnamefont {M.~A.}\ \bibnamefont
  {Castellanos-Beltran}}, \bibinfo {author} {\bibfnamefont {K.~D.}\
  \bibnamefont {Irwin}}, \bibinfo {author} {\bibfnamefont {G.~C.}\ \bibnamefont
  {Hilton}}, \bibinfo {author} {\bibfnamefont {L.~R.}\ \bibnamefont {Vale}}, \
  and\ \bibinfo {author} {\bibfnamefont {K.~W.}\ \bibnamefont {Lehnert}},\
  }\href@noop {} {\bibfield  {journal} {\bibinfo  {journal} {Nat. Phys.}\
  }\textbf {\bibinfo {volume} {4}},\ \bibinfo {pages} {928} (\bibinfo {year}
  {2008})}\BibitemShut {NoStop}%
\bibitem [{\citenamefont {Bergeal}\ \emph
  {et~al.}(2010{\natexlab{b}})\citenamefont {Bergeal}, \citenamefont
  {Schackert}, \citenamefont {Metcalfe}, \citenamefont {Vijay}, \citenamefont
  {Manucharyan}, \citenamefont {Frunzio}, \citenamefont {Prober}, \citenamefont
  {Schoelkopf}, \citenamefont {Girvin},\ and\ \citenamefont
  {Devoret}}]{Bergeal_Nature2010}%
  \BibitemOpen
  \bibfield  {author} {\bibinfo {author} {\bibfnamefont {N.}~\bibnamefont
  {Bergeal}}, \bibinfo {author} {\bibfnamefont {F.}~\bibnamefont {Schackert}},
  \bibinfo {author} {\bibfnamefont {M.}~\bibnamefont {Metcalfe}}, \bibinfo
  {author} {\bibfnamefont {R.}~\bibnamefont {Vijay}}, \bibinfo {author}
  {\bibfnamefont {V.~E.}\ \bibnamefont {Manucharyan}}, \bibinfo {author}
  {\bibfnamefont {L.}~\bibnamefont {Frunzio}}, \bibinfo {author} {\bibfnamefont
  {D.~E.}\ \bibnamefont {Prober}}, \bibinfo {author} {\bibfnamefont {R.~J.}\
  \bibnamefont {Schoelkopf}}, \bibinfo {author} {\bibfnamefont {S.~M.}\
  \bibnamefont {Girvin}}, \ and\ \bibinfo {author} {\bibfnamefont {M.~H.}\
  \bibnamefont {Devoret}},\ }\href {http://dx.doi.org/10.1038/nature09035}
  {\bibfield  {journal} {\bibinfo  {journal} {Nature}\ }\textbf {\bibinfo
  {volume} {465}},\ \bibinfo {pages} {64} (\bibinfo {year}
  {2010}{\natexlab{b}})}\BibitemShut {NoStop}%
\bibitem [{\citenamefont {Hatridge}\ \emph {et~al.}(2011)\citenamefont
  {Hatridge}, \citenamefont {Vijay}, \citenamefont {Slichter}, \citenamefont
  {Clarke},\ and\ \citenamefont {Siddiqi}}]{Hatridge_PRB2011}%
  \BibitemOpen
  \bibfield  {author} {\bibinfo {author} {\bibfnamefont {M.}~\bibnamefont
  {Hatridge}}, \bibinfo {author} {\bibfnamefont {R.}~\bibnamefont {Vijay}},
  \bibinfo {author} {\bibfnamefont {D.~H.}\ \bibnamefont {Slichter}}, \bibinfo
  {author} {\bibfnamefont {J.}~\bibnamefont {Clarke}}, \ and\ \bibinfo {author}
  {\bibfnamefont {I.}~\bibnamefont {Siddiqi}},\ }\href {\doibase
  10.1103/PhysRevB.83.134501} {\bibfield  {journal} {\bibinfo  {journal} {Phys.
  Rev. B}\ }\textbf {\bibinfo {volume} {83}},\ \bibinfo {pages} {134501}
  (\bibinfo {year} {2011})}\BibitemShut {NoStop}%
\bibitem [{\citenamefont {Roch}\ \emph {et~al.}(2012)\citenamefont {Roch},
  \citenamefont {Flurin}, \citenamefont {Nguyen}, \citenamefont {Morfin},
  \citenamefont {Campagne-Ibarcq}, \citenamefont {Devoret},\ and\ \citenamefont
  {Huard}}]{Roch_PRL2012}%
  \BibitemOpen
  \bibfield  {author} {\bibinfo {author} {\bibfnamefont {N.}~\bibnamefont
  {Roch}}, \bibinfo {author} {\bibfnamefont {E.}~\bibnamefont {Flurin}},
  \bibinfo {author} {\bibfnamefont {F.}~\bibnamefont {Nguyen}}, \bibinfo
  {author} {\bibfnamefont {P.}~\bibnamefont {Morfin}}, \bibinfo {author}
  {\bibfnamefont {P.}~\bibnamefont {Campagne-Ibarcq}}, \bibinfo {author}
  {\bibfnamefont {M.~H.}\ \bibnamefont {Devoret}}, \ and\ \bibinfo {author}
  {\bibfnamefont {B.}~\bibnamefont {Huard}},\ }\href {\doibase
  10.1103/PhysRevLett.108.147701} {\bibfield  {journal} {\bibinfo  {journal}
  {Phys. Rev. Lett.}\ }\textbf {\bibinfo {volume} {108}},\ \bibinfo {pages}
  {147701} (\bibinfo {year} {2012})}\BibitemShut {NoStop}%
\bibitem [{\citenamefont {Hatridge}\ \emph {et~al.}(2012)\citenamefont
  {Hatridge}, \citenamefont {Shankar}, \citenamefont {Mirrahimi}, \citenamefont
  {Schackert}, \citenamefont {Geerlings}, \citenamefont {Brecht}, \citenamefont
  {Sliwa}, \citenamefont {Abdo}, \citenamefont {Frunzio}, \citenamefont
  {Girvin}, \citenamefont {Schoelkopf},\ and\ \citenamefont
  {Devoret}}]{Hatridge_Science2012}%
  \BibitemOpen
  \bibfield  {author} {\bibinfo {author} {\bibfnamefont {M.}~\bibnamefont
  {Hatridge}}, \bibinfo {author} {\bibfnamefont {S.}~\bibnamefont {Shankar}},
  \bibinfo {author} {\bibfnamefont {M.}~\bibnamefont {Mirrahimi}}, \bibinfo
  {author} {\bibfnamefont {F.}~\bibnamefont {Schackert}}, \bibinfo {author}
  {\bibfnamefont {K.}~\bibnamefont {Geerlings}}, \bibinfo {author}
  {\bibfnamefont {T.}~\bibnamefont {Brecht}}, \bibinfo {author} {\bibfnamefont
  {K.~M.}\ \bibnamefont {Sliwa}}, \bibinfo {author} {\bibfnamefont
  {B.}~\bibnamefont {Abdo}}, \bibinfo {author} {\bibfnamefont {L.}~\bibnamefont
  {Frunzio}}, \bibinfo {author} {\bibfnamefont {S.~M.}\ \bibnamefont {Girvin}},
  \bibinfo {author} {\bibfnamefont {R.~J.}\ \bibnamefont {Schoelkopf}}, \ and\
  \bibinfo {author} {\bibfnamefont {M.~H.}\ \bibnamefont {Devoret}},\
  }\href@noop {} {\bibfield  {journal} {\bibinfo  {journal} {Accepted for
  Publication in Science}\ } (\bibinfo {year} {2012})}\BibitemShut {NoStop}%
\bibitem [{\citenamefont {Abdo}\ \emph {et~al.}(2011)\citenamefont {Abdo},
  \citenamefont {Schackert}, \citenamefont {Hatridge}, \citenamefont
  {Rigetti},\ and\ \citenamefont {Devoret}}]{Abdo_APL2011}%
  \BibitemOpen
  \bibfield  {author} {\bibinfo {author} {\bibfnamefont {B.}~\bibnamefont
  {Abdo}}, \bibinfo {author} {\bibfnamefont {F.}~\bibnamefont {Schackert}},
  \bibinfo {author} {\bibfnamefont {M.}~\bibnamefont {Hatridge}}, \bibinfo
  {author} {\bibfnamefont {C.}~\bibnamefont {Rigetti}}, \ and\ \bibinfo
  {author} {\bibfnamefont {M.}~\bibnamefont {Devoret}},\ }\href {\doibase
  10.1063/1.3653473} {\bibfield  {journal} {\bibinfo  {journal} {Applied
  Physics Letters}\ }\textbf {\bibinfo {volume} {99}},\ \bibinfo {eid} {162506}
  (\bibinfo {year} {2011})}\BibitemShut {NoStop}%
\bibitem [{\citenamefont {Abdo}\ \emph
  {et~al.}(2012{\natexlab{b}})\citenamefont {Abdo}, \citenamefont {Sliwa},
  \citenamefont {Schackert}, \citenamefont {Bergeal}, \citenamefont {Hatridge},
  \citenamefont {Frunzio}, \citenamefont {Stone},\ and\ \citenamefont
  {Devoret}}]{Abdo_InPreparation2012}%
  \BibitemOpen
  \bibfield  {author} {\bibinfo {author} {\bibfnamefont {B.}~\bibnamefont
  {Abdo}}, \bibinfo {author} {\bibfnamefont {K.}~\bibnamefont {Sliwa}},
  \bibinfo {author} {\bibfnamefont {F.}~\bibnamefont {Schackert}}, \bibinfo
  {author} {\bibfnamefont {N.}~\bibnamefont {Bergeal}}, \bibinfo {author}
  {\bibfnamefont {M.}~\bibnamefont {Hatridge}}, \bibinfo {author}
  {\bibfnamefont {L.}~\bibnamefont {Frunzio}}, \bibinfo {author} {\bibfnamefont
  {A.~D.}\ \bibnamefont {Stone}}, \ and\ \bibinfo {author} {\bibfnamefont
  {M.~H.}\ \bibnamefont {Devoret}},\ }\href@noop {} {\bibfield  {journal}
  {\bibinfo  {journal} {In Preparation}\ } (\bibinfo {year}
  {2012}{\natexlab{b}})}\BibitemShut {NoStop}%
\bibitem [{\citenamefont {Chong}\ \emph {et~al.}(2010)\citenamefont {Chong},
  \citenamefont {Ge}, \citenamefont {Cao},\ and\ \citenamefont
  {Stone}}]{Chong_PRL2010}%
  \BibitemOpen
  \bibfield  {author} {\bibinfo {author} {\bibfnamefont {Y.~D.}\ \bibnamefont
  {Chong}}, \bibinfo {author} {\bibfnamefont {L.}~\bibnamefont {Ge}}, \bibinfo
  {author} {\bibfnamefont {H.}~\bibnamefont {Cao}}, \ and\ \bibinfo {author}
  {\bibfnamefont {A.~D.}\ \bibnamefont {Stone}},\ }\href {\doibase
  10.1103/PhysRevLett.105.053901} {\bibfield  {journal} {\bibinfo  {journal}
  {Phys. Rev. Lett.}\ }\textbf {\bibinfo {volume} {105}},\ \bibinfo {pages}
  {053901} (\bibinfo {year} {2010})}\BibitemShut {NoStop}%
\bibitem [{\citenamefont {Wan}\ \emph {et~al.}(2011)\citenamefont {Wan},
  \citenamefont {Chong}, \citenamefont {Ge}, \citenamefont {Noh}, \citenamefont
  {Stone},\ and\ \citenamefont {Cao}}]{Wan_Science2011}%
  \BibitemOpen
  \bibfield  {author} {\bibinfo {author} {\bibfnamefont {W.}~\bibnamefont
  {Wan}}, \bibinfo {author} {\bibfnamefont {Y.}~\bibnamefont {Chong}}, \bibinfo
  {author} {\bibfnamefont {L.}~\bibnamefont {Ge}}, \bibinfo {author}
  {\bibfnamefont {H.}~\bibnamefont {Noh}}, \bibinfo {author} {\bibfnamefont
  {A.~D.}\ \bibnamefont {Stone}}, \ and\ \bibinfo {author} {\bibfnamefont
  {H.}~\bibnamefont {Cao}},\ }\href {\doibase 10.1126/science.1200735}
  {\bibfield  {journal} {\bibinfo  {journal} {Science}\ }\textbf {\bibinfo
  {volume} {331}},\ \bibinfo {pages} {889} (\bibinfo {year}
  {2011})}\BibitemShut {NoStop}%
\bibitem [{Note1()}]{Note1}%
  \BibitemOpen
  \bibinfo {note} {We chose to present these two data sets as this is where we
  best achieved the balance between signal and idler photon flux.}\BibitemShut
  {Stop}%
\bibitem [{\citenamefont {Pozar}(1997)}]{Pozar_Book1997}%
  \BibitemOpen
  \bibfield  {author} {\bibinfo {author} {\bibfnamefont {D.}~\bibnamefont
  {Pozar}},\ }\href {http://books.google.com/books?id=IDxTAAAAMAAJ} {\emph
  {\bibinfo {title} {Microwave engineering}}}\ (\bibinfo  {publisher} {Wiley},\
  \bibinfo {year} {1997})\ p.\ \bibinfo {pages} {549}\BibitemShut {NoStop}%
\bibitem [{\citenamefont {Dicke}(1946)}]{Dicke_RevSciInstrum1946}%
  \BibitemOpen
  \bibfield  {author} {\bibinfo {author} {\bibfnamefont {R.~H.}\ \bibnamefont
  {Dicke}},\ }\href {\doibase 10.1063/1.1770483} {\bibfield  {journal}
  {\bibinfo  {journal} {Review of Scientific Instruments}\ }\textbf {\bibinfo
  {volume} {17}},\ \bibinfo {pages} {268} (\bibinfo {year} {1946})}\BibitemShut
  {NoStop}%
\end{thebibliography}
\bibliographystyle{apsrev4-1}
%\bibliographystyle{apsrev}
%%%%%%%%%%%%%%%%%%%%%%%%%%%%%%%%%%%%%%%%%%%%%%%%%%%%%%%%%%%%%%%%%%%%%%%%%%%%%%%%%%%%%%%%%%%%%%%%%
%%%%%%%%%% BIBLIOGRAPHY: copied from latex generated .bbl file

%%%%%%%%%%%%%%%%%%%%%%%%%%%%%%%%%%%%%%%%%%%%%%%%%%%%%%%%%%%%%%%%%%%%%%%%%%%%%%%%%%%%%%%%%%%%%%%%%
\end{document}